\documentstyle{article}
\begin{document}
\rightline{May 1993}
\rightline{McGill/93-32}

\vskip 1.2cm
\begin{center}
\begin{large}
{\bf Electric charge quantization without anomalies?}
\vskip 1.5cm
\end{large}
R. Foot
\end{center}
\vskip 1cm
\noindent
Physics Department, McGill university, 3600 University
street, Montreal, Quebec, H3A 2T8,
Canada.
\vskip 2cm
\begin{center}
{\bf Abstract}
\end{center}
\vskip 1cm
In gauge theories like the standard model, the electric charges of the
fermions can be heavily constrained from the classical structure of
the theory and from the cancellation of anomalies.
We argue that the anomaly conditions are not quite as well motivated
as the classical constraints, since it is possible that
new fermions could exist which cancel potential anomalies.
For this reason we examine the classically
allowed electric charges of the known fermions and we point out
that the electric charge of the tau neutrino is classically allowed
to be non-zero. The experimental bound on the electric charge of the
tau neutrino is many orders of magnitude weaker than for any other
known neutrino.
We discuss possible modifications of the minimal standard model
such that electric charge is quantized classically.

\newpage

The quantization of the electric charges of most of the known fermions
is a well established experimental phenomenon. An approach to
a theoretical understanding of this phenomenon
has emerged in recent years based on the standard model [1].
The standard model is a gauge theory with gauge group
$$SU(3)_c \otimes SU(2)_L \otimes U(1)_Y,    \eqno (1)$$
which is assumed to be spontaneously broken by the vacuum
expectation value (VEV) of a scalar doublet $\phi \sim (1, 2, 1)$
(whose $U(1)_Y$ charge can be normalized to $1$ without loss of generality due
to a scaling symmetry; $g \rightarrow \eta g, Y \rightarrow Y/\eta$,
where $g$ is the $U(1)_Y$ coupling constant, and $Y$ is the generator of the
$U(1)_Y$ gauge group.). The gauge symmetry of the Lagrangian can be used to
choose the standard form for the vacuum:
$$\langle \phi \rangle = \left(\begin{array}{c}
0\\
u
\end{array}\right). \eqno (2)$$
The VEV of $\phi$ breaks $SU(2)_L \otimes U(1)_Y$ (but does'nt break $SU(3)_c$
of course) leaving
an unbroken $U(1)$ symmetry, $U(1)_Q$, which is identified with
electromagnetism and its generator $Q$
is the linear combination (which annihilates the VEV of eq.(2)):
$$Q = I_3 + Y/2. \eqno(3)$$
(The normalization of $Q$ is undetermined, and we have taken the convention
of normalizing it so that the charged $W$ bosons will have charge 1.)

There are two quite distinct ways in which the standard model
constrains the electric charges of the fermions.
Firstly, there are a set of constraints which follow from
the consistency of the theory at the classical level
(such as the requirement
that the Lagrangian be gauge invariant), while there are other
constraints which follow from the consistency of the theory at the
quantum level (i.e. the anomaly cancellation conditions).
We first discuss the classical constraints.
The invariance of the Yukawa Lagrangian (or equivalently,
the electromagnetic invariance of the
fermion mass terms and the
quark flavour mixing terms in the weak interaction)
constrains the electric charges.
For example, the electron mass term in the lagrangian
is invariant under $U(1)_Q$ if and only if
the electric charge of the left-handed electron is equal to the charge of
the right-handed electron. A similiar arguement holds for the quarks,
so that for one generation, there are four electric
charges, the charge of the electron, the charge of the neutrino,
the charge of the up quark and the charge of the down quark [2].
A second piece of information about the fermion electric charges can
be obtained by observing that the left-handed fermions are in $SU(2)_L$
doublets. Since $SU(2)_L$ and $U(1)_Y$ are a direct product (i.e act
independently of each other), the members of the $SU(2)_L$ doublet
have the same $U(1)_Y$ charge and thus the difference of
the electric charges of the members of the $SU(2)_L$ doublet
is just the difference of their $I_3$ eigenvalue (which is just equal
to 1 with our normalization). Hence we have the information that
the electric charge of the electron neutrino minus the electric
charge of the electron is 1,
and the electric charge of the up quark minus the electric
charge of the down quark
is equal to 1. So in each generation there are only two unknown electric
charges,
which can be taken as the electric
charge of the neutrino and the electric charge of the down quark.

Since the CKM matrix is non-diagonal, there are additional classical
constraints from the flavour mixing terms in the standard model Lagrangian.
For example, the $W$ boson couples a $u$ quark to a $s$ quark, as
well as a $u$ quark to a $d$ quark. Since the Lagrangian is invariant
under $U(1)_Q$, the existence of these terms tells us that the $s$ and $d$
quark electric charges are equal. A similiar mixing happens of course with the
third generation quarks, so that the electric charge of the $b$ quark
must be equal to the electric charges of the $s$ and $d$ quarks.
Hence each of the three known generations of quarks have exactly
the same electric charges. No such mixing has been observed in
the lepton sector, and in the minimal standard model there can
be no such mixing as the neutrinos are massless in that model.
Hence the constraints from mass and mixing together with the $SU(2)_L$ doublet
structure of the left-handed fermions tell us
that there are {\it four} classically undetermined
electric charges in the standard model. These four undertermined electric
charges can be taken to be the electric
charges of the three neutrinos
and the down quark (we denote these four electric charges as
$Q(\nu_e), Q(\nu_{\mu}), Q(\nu_{\tau}),$ and $Q(d)$).
All of the other fermion electric charges can be
uniquely determined in terms of these four classically undetermined parameters.
Experimentally, it is known that
\begin{eqnarray}
&Q(d)& = -1/3 \pm \delta_d, \    \delta_d < 10^{-21} \nonumber \\
&Q(\nu_e)& = 0 \pm \delta_{\nu_e}, \  \delta_{\nu_e} <  10^{-21}\nonumber \\
&Q(\nu_{\mu})& = 0 \pm \delta_{\nu_{\mu}}, \   \delta_{\nu_{\mu}} <  10^{-9}
\nonumber \\
&Q(\nu_{\tau})& = 0 \pm \delta_{\nu_{\tau}}, \
\delta_{\nu_{\tau}} <\ 3 \times 10^{-4} \nonumber
\end{eqnarray}
$$\eqno (4)$$
where the experimental bounds (i.e the delta parameters) come from
experiments on the neutron charge [3], experiments on the neutrality of
matter [4], experiments on $\nu_{\mu} e$ scattering [5].
The experimental bound on the electric charge of the
$\nu_{\tau}$ has not been specifically studied previously (as far as
we are aware) and the constraint given in Eq.(4) comes from an
analysis in Ref.[6] which examines the experimental bounds on the
electric charge of a  hypothetical
exotic ``mini-charged'' particle [7].

At this stage one can argue that further constraints can
be obtained by assuming that gauge anomalies cancel. Anomalies
imply the loss of a classical symmetry in the quantum theory [8].
If we assume that triangle anomalies cancel then
we have two constraints which are not independent of the classical constraints.
They are the $[U(1)_Q]^3$ and $[SU(2)]^2 U(1)_Q$ anomaly condition.
The cancellation of the $[U(1)_Q]^3$ anomaly gives the constraint:
$$ Q(\nu_e)^3 + Q(\nu_{\mu})^3 + Q(\nu_{\tau})^3 = 0. \eqno (5)$$
The equation only involves the neutrinos, since there is no contribution
from the charged fermions (this is because the classical constraints
derived from the existence of non-zero masses for the charged fermions
implies that $U(1)_Q$ is vector like for the charged fermions).
The $[SU(2)_L]^2U(1)_Q$ anomaly cancellation condition implies that
$$Q(\nu_e) + Q(\nu_{\mu}) + Q(\nu_{\tau}) + 9Q(d) = -3. \eqno (6)$$
Thus there are now only two undermined electric charges.
One further independent equation can be obtained from the mixed
gauge-gravitational anomaly cancellation [9], which says that:
$$Q(\nu_e) + Q(\nu_{\mu}) + Q(\nu_{\tau}) = 0. \eqno (7)$$
Thus we are left with one undetermined electric charge, which it turns out
must be taken as a lepton charge (since eq.(7) and (6) uniquely determine
$Q(d) = -1/3$ and hence all the quark charges have been determined).
Thus one must conclude that the minimal standard model does not have
electric charge quantization. There is one free parameter.
Thus an understanding of electric charge quantization
requires new physics beyond the minimal standard model.
Various ways of extending the standard model so that electric
charge is quantized have been discussed in the liturature [1].
One can simply add some terms to the standard model Lagrangian which
yield additional constraints. Perhaps the most obvious (and also well
motivated) way to do this is
to add neutrino masses.
For example, one can add three right-handed gauge singlet neutrinos
with Dirac and Majorana mass terms. One can assume that there is
a non-diagonal CKM type matrix for the leptons which will imply that
each generation of leptons have equal charges (just like in the case
of the quarks). In addition, Majorana masses for the right-handed
neutrinos will fix the charge of the neutrinos to zero.
This extension of the standard model would then have every electric
charge (ratio) completely determined and hence electric charge
quantization would be understood in terms of the internal consistency of
the theory.

There is one important point that should be mentioned.
The quantum constraints which are the anomaly cancellation
equations are not quite as well motivated as the classical
constraints.
For example, all of the classical constraints seem to be very strong
constraints in the sense that we know for certain that the electron
has a mass. We know for certain that a coupling of $W$ to $u$ and $d$ and
$u$ and $s$ exists etc.  Therefore, under the assumption that electric
charge is conserved, our conclusions derived from the
classical structure of the theory, such as $Q(u) - Q(d) = Q(\nu_e) - Q(e) = 1$
and $Q(d) = Q(s) = Q(b)$ seems to be unchallengable.
On the other hand, the anomaly constraints are not definitely true.
For example, there could exist a set of ``mirror'' fermions which
have the same gauge quantum numbers as the standard fermions (except
that left and right chiralities are interchanged), but
are too heavy to be seen yet in experiment. In this case, there would be
{\it no} nontrivial anomaly cancellation equations.
So, we feel that it is an interesting question as to whether gauge theories
with $U(1)$ factors, can have electric charge quantization classically i.e.
can the classical constraints be sufficient to determine all of the
electric charges?

So, if we ignore the constraints from anomalies, then as discussed
above there are four classically undertermined parameters in the minimal
standard model. One can see from Eq.(4) that two of these
parameters are extremely well constrained (to within $10^{-21}$), one
of them is moderately well constrained (to within $10^{-9}$) and
one of them, the electric charge of the tau neutrino
is not well constrained [10] (note however that there are
significant indirect bounds on the electric charge from astrophysics
if the tau neutrino has a mass less than about 25 keV [7]).
Since it is theoretically possible for the
charge of the tau neutrino to be non-zero and since as far as we
are aware, a non-zero tau neutrino electric charge has never
been searched for in experiments, we propose such an experiment to
put to the test the standard assumption that the tau neutrino is neutral.

Of course the minimal standard model may not be complete. It is interesting
to look for ways to modify the model so that electric charge is quantized
classically.
If right-handed neutrinos exist and they have Dirac mass terms with the usual
left-handed neutrinos and we assume that nontrivial mixing effects in the weak
interaction occur (just like in the quark sector),
then in this case, $Q(\nu_e) = Q(\nu_{\mu}) = Q(\nu_{\tau})$,
so that there are only two classically undetermined electric charges,
which can be taken to be $Q(\nu_e)$ and $Q(d)$.
If there is a Majorana mass term for one or more of the right-handed neutrinos
then one obtains the additional constraint that $Q(\nu_e) = 0$.
Thus, in this case, there is only one undetermined electric charge, which can
be taken to be the electric charge of the down quark, $Q(d)$.

Following our philosophy, we need to modify the Lagrangian so that
$Q(d)$ is uniquely determined. Another way of thinking about
the problem is in terms of global $U(1)$ symmetries.
At the classical level, the minimal standard model Lagrangian has four
global symmetries: $U(1)_{L_e}, U(1)_{L_{\mu}}, U(1)_{L_{\tau}}, U(1)_B$
and one local symmetry $U(1)_Y$.
At the classical level, there is no theoretical reason why any combination
of $Y$ and $L_e, L_{\mu}, L_{\tau}, B$ cannot be the one $U(1)$ which
is gauged.
This means that there is a four parameter uncertainty in the $U(1)$ which
is gauged.
When we modify the lepton sector by adding right-handed (gauge singlet)
neutrinos and include mass and mixing terms for the neutrinos,
then this new Lagrangian has, in general, only one global
symmetry, which is baryon number $U(1)_B$.
Hence at the classical level, any combination of $Y$ and $B$ is
a $U(1)$ symmetry and can be the $U(1)$ which is gauged.
To obtain correct electric charge quantization, we must
modify the theory such that baryon number is violated (but
with $Y$ left conserved of course).
Unlike the case of the lepton sector, we cannot do this by simply
adding Majorana mass terms. This works for the leptons since
Majorana masses violate the global lepton number (but conserve
standard hypercharge).
However for quarks, any Majorana mass would violate both baryon number
and standard hypercharge (leaving some linear combination conserved which would
consequently not correspond to the electric charges of the real world).
Assuming only the standard model gauge symmetry, then the simplest
way that we know about to modify the theory to obtain electric charge
quantization is to add a new scalar such that its interactions violate
the baryon number symmetry (but conserve hypercharge) [11]. The scalar
must interact with quarks if it is to violate baryon number.
Assuming the usual renormalizable dimension four (Yukawa-type) coupling,
then there are
only a finite number of possible quantum numbers
for the scalar.
Since the scalar will couple to a fermion bilinear, it follows
from gauge invariance that the quantum numbers of the
scalar are those of the fermion bilinears.
For example, a scalar $\sigma_1$ coupling via the interaction term
${\cal L} = \lambda \sigma_1^{\dagger} \bar Q_L (f_L)^c$ implies
that $\sigma_1$ transforms like $\bar Q_L (f_L)^c$.
Thus we can simply list the possible scalars in terms of fermion
bilinears with $SU(3)_c \otimes SU(2)_L \otimes U(1)_Y$
representations as follows:
\begin{eqnarray}
&\sigma_1& \sim \bar Q_L (f_L)^c \sim (\bar 3, 1 + 3, -y_d) \nonumber \\
&\sigma_2& \sim \bar Q_L e_R \sim \bar u_R f_L \sim (\bar 3, 2, -3-y_d)
\nonumber \\
&\sigma_3& \sim \bar Q_L (Q_L)^c \sim (3 + \bar 6, 1 + 3, -2 - 2y_d)
\nonumber \\
&\sigma_4& \sim \bar u_R (d_R)^c \sim (3 + \bar 6, 1, -2-2y_d) \nonumber \\
&\sigma_5& \sim \bar u_R (e_R)^c \sim \bar d_R (\nu_R)^c
\sim (\bar 3, 1, - y_d) \nonumber \\
&\sigma_6& \sim \bar u_R (\nu_R)^c \sim (\bar 3, 1, -2 - y_d) \nonumber \\
&\sigma_7& \sim \bar d_R f_L \sim \bar Q_L \nu_R \sim (\bar 3, 2, -1-y_d)
\nonumber \\
&\sigma_8& \sim \bar u_R (u_R)^c \sim (3 + \bar 6, 1, -4-2y_d)
\nonumber\\
&\sigma_9& \sim \bar d_R (d_R)^c \sim (3 + \bar 6, 1, -2y_d)
\nonumber
\end{eqnarray}
$$\eqno (8)$$
where our notation for the standard model fermions
(+ right-handed neutrinos) is as follows:
\begin{eqnarray}
&f_L \sim (1, 2, -1),\quad  e_R \sim (1, 1, -2), \quad \nu_R \sim (1, 1, 0),
& \nonumber \\
&Q_L \sim (3, 2, 1 + y_d), \quad u_R \sim (3, 1, 2 + y_d), \quad d_R
\sim (3, 1, y_d), & \nonumber
\end{eqnarray}
$$\eqno (9)$$
with generation index suppressed.
We will assume for simplicity that there exists only one exotic
scalar. Note that the above interactions do not, by themselves break
baryon number since the scalar can carry baryon number.
We need to break baryon number in the scalar potential.
Note that since all of the scalars are either in the $3$ or $6$
representation of $SU(3)_c$, the smallest dimensional term which
breaks baryon number and conserves $SU(3)_c$ is the trilinear term
$\sigma^3$.
(Note that there is no quadratic or quatic term which breaks baryon number and
conserves $SU(3)_c$).
Any $\sigma^3$ term will also violate standard hypercharge.
The only possible renormalizable term
must involve 3 sigma's and the Higgs doublet $\phi$.
Since the Higgs doublet has hypercharge 1 (in our normalization),
a $\sigma^3 \phi$ or $\sigma^3 \phi^{\dagger}$ term
will imply that the $\sigma$ scalar must have hypercharge $-1/3$ or $1/3$
respectively.
The only candidate for $\sigma$ is $\sigma_7$ since any other choice
will clearly lead to the wrong hypercharge assignments.
[For example a $\sigma_1^3 \phi$
would constrain the hypercharge of $\sigma_1$ to be $-1/3$,
which will consequently constrain $y_d = -1/3$. Then using eq.(3),
we would find that the electric charge of the $d$ quark would by $-1/6$
which would lead to incorrect electric charges for the hadrons, and
of course does not correspond to the real world.]
Thus, we conclude that under the assumption of only one
exotic scalar, electric charge can be quantized classically.
Furthermore the quantum numbers of the scalar and the form
of the interactions of the scalar are uniquely determined.
The scalar couples leptons to quarks
through the Lagrangian terms:
$${\cal L} = \lambda_1 \bar f_L \sigma d_R + \lambda_2 \bar Q_L \sigma^c
\nu_R +
H.c., \eqno (10)$$
where gauge invariance of this Lagrangian term implies that
$$\sigma \sim (\bar 3, 2, -y_d -1). \eqno (11)$$
The hypercharge of $\sigma$
is constrained to be $-1/3$ (which means that $y_d$ is constrained to be
$-2/3$) by the scalar potential terms:
$$\Delta V(\phi, \sigma) = \lambda \sigma^3 \phi + H.c. \eqno (12)$$
Thus, the interactions of $\sigma$ fix the undertermined
hypercharge of the $d$ quark, resulting in a model with
electric charge quantization at the classical level.
Note that we must choose the parameters in the scalar potential
such that $\sigma$ does not get any VEV,
while $\phi$ of course gets a VEV. It is straightforward
to show that this is possible.
We leave the details as an exercise to the reader.

Since $\sigma$ violates baryon number, interactions
involving $\sigma$ will induce baryon number violating processes.
The process which should place the most stringent limit on
the mass of $\sigma$ will be the experimental bound on proton decay.
Observe that any Feynman diagram leading to proton decay
must involve the baryon number violating $\sigma^3\phi$
interaction. The leading order diagram for the proton decay
involves one of these interactions, and thus contains three $\sigma$
fields (note that when the VEV of $\phi$ is included,
the $\sigma^3 \phi$ interaction containes a $\sigma^3$ interaction
term). One can easily see that the simplest diagram giving proton decay
leads to the decay $P \rightarrow \pi^+ + \nu + \nu + \nu$. The
order of magnitude of the decay width
for this decay can be evaluated from
simple dimensional arguements,
$$\Gamma(P\rightarrow \pi^+ + \nu + \nu + \nu) \sim
{\cal O}\left({\langle \phi \rangle^2 M_P^{11}
\over M_{\sigma}^{12}}\right), \eqno (13)$$
where $M_P$ is the proton mass, and $M_{\sigma}$ is the $\sigma$
scalar mass.
Thus, applying the existing experimental limit on the lifetime
of the proton
we find that the mass of $\sigma$ is constrained to be
greater than about $10^5$ GeV.

Finally note that this model may be easily modified so that
only the electric charge parameter $Q(\nu_{\tau})$ is undetermined.
The $\sigma$ field can be introduced with interactions described
above to fix $Q(d) = -1/3$. In the lepton sector it is possible
that the third generation does not mix with the first two generations
and that the third generation neutrino (i.e. the tau neutrino) is a Dirac
fermion (note that a Majorana fermion must have zero electric charge
if electromagnetism is unbroken).
In this case, the resulting model
would have $Q(\nu_{\tau})$ classically undertermined.
Its mass can be large enough (i.e. greater than about 25 keV [7])
to evaide the astrophysical constraints.
Thus, we emphasise again the importance of putting the standard
assumption of a electrically neutral $\nu_{\tau}$ to the test.

For completeness we mention that a different type of mechanism
for obtaining electric charge quantization in a theory with
a $U(1)$ gauge factor is possible if the gauge group is
enlarged so that a discrete symmetry interchanging the quarks and leptons
is assumed. The resulting quark-lepton symmetric models can have
a $U(1)$ factor in the gauge group which can be
completely fixed classically [12].

In conclusion, we have discussed the issue of electric charge quantization
in the standard model.
There are two different ways in which the minimal standard model
constrains the electric charges of the fermions.
There are constraints which follow from the classical structure of the
theory and those which follow from the quantization of the theory
(i.e. anomaly cancellation). We argue that the classical constraints are
very well motivated constraints, while the anomaly conditions are not as
well motivated, since new (heavy) fermions could exist which cancel
any potential anomalies. We examined the classically allowed electric
charges of the fermions in the minimal standard model.
We made the observation that the electric charge of
the $\tau$ neutrino may be non-zero, and that the current experimental
constraints on the charge of the $\tau$ neutrino seem to be very weak.
In fact there may be no experimental searches for a charged
$\tau$ neutrino at all. If this is the case, then we argue that
experiments should be undertaken to test the neutrality of the tau neutrino.
We then examined ways in which the standard model could be modified so that
the electric charges are quantized correctly classically.

\vskip 1cm
\noindent
{\bf References:}
\vskip .5cm
\noindent
[1]  For a review, see R. Foot, G. C. Joshi, H. Lew, and R. R. Volkas,
Mod. Phys. Lett. A5, 2721 (1990).
\vskip .5cm
\noindent
[2] Note that we have implicitly assumed that $SU(3)_c$ is unbroken and hence
each quark colour will have the same electric charge. Thus, there are only
two quark charges in each generation.
\vskip .5cm
\noindent
[3] J. Baumann, J. Kalus, R. Gahler, and W. Mampe, Phys. Rev. D37, 3107
(1988). This paper gives a experimental limit on the neutron charge.
This translates into a limit on the down quark charge since,
$Q(N) = Q(u) + 2Q(d) = 3Q(d) + 1$
\vskip 0.5cm
\noindent
[4] M. Marinelli and G. Morpugo, Phys. Lett. 137B, 439 (1984);
see also H. F. Dylla and J. G. King, Phys. Rev. A7, 1224 (1973).
These papers give the limit on the neutrality of matter.
This limit, together with the limit on the down quark charge
obtained from Ref [3], gives a limit on the electron neutrino charge
(assuming the classical constraint: $Q(\nu_e) - Q(e) = Q(u) - Q(d)$).
\vskip 0.5cm
\noindent
[5] K. S. Babu and R. R. Volkas, Phys. Rev. D42, 2764 (1992).
\vskip 0.5cm
\noindent
[6] S. Davidson, B. Campbell and D. Bailey, Phys. Rev. D43, 2314 (1991).
Note that M. I. Dobroliubov and A. Yu. Ignative [Phys. Rev. Lett.6, 679 (1990)]
have also considered bounds on the electric charge of a hypothetical
mini-charged
particle but their experimental bounds seem to be weaker than the
$3 \times 10^{-4}$ bound of Davidson et al.
\vskip 0.5cm
\noindent
[7] In addition to direct experimental bounds obtained from Laboratory
experiments, there are in the liturature indirect bounds on the charge of
a hypothetical mini-charged particle. In particular, an upper limit of
$10^{-13}e$ can be obtained from requiring that plasmon decay into
neutrino pairs in red giants be not too efficient
[see J. Berstein et al, Phys. Rev.  1963]. This bound is applicable only
when the charged particle has mass less than 25 keV. So, if this
bound is taken seriously, a tau neutrino can only have a experimentally
interesting electric charge if its mass is greater than 25 keV.
Also note that an electric charge in the range $10^{-9} \le Q(\nu_{\tau}) \le
10^{-7}$ has been shown to be inconsistent with observations of
SN1987A [see R. N. Mohapatra and I. Z. Rothstein, Phys.Lett.B247, 593 (1990);
R. N. Mohapatra and S.N. Nussinov, Int. J. Mod. Phys. A8, 3817 (1992).]
\vskip 0.5cm
\noindent
[8] S. Adler, Phys. Rev. 177, 2426 (1969); J. S. Bell and R. Jackiw,
Nuovo Cimento 60A, 49 (1969).
\vskip 0.5cm
\noindent
[9] R. Delbourgo and A. Salam, Phys. Lett. 40B, 381 (1972);
T. Eguchi and P. Freund, Phys. Rev. Lett. 37, 1251 (1976);
L. Alvarez-Gaume and E. Witten, Nucl. Phys. B234, 269 (1983).
\vskip .5cm
\noindent
[10] Note that a non-zero value for $Q(\nu_{\tau})$ will imply that
the charge of the tau ($Q(\tau) = 1 + Q(\nu_{\tau})$) will be slightly
different to the electron electric charge. However, we expect that the effect
of a non-zero tau neutrino charge should be the most important
phenomenologically.
\vskip .5cm
\noindent
[11] Note that the resulting model was mentioned by a footnote in Foot,
Joshi, Lew and Volkas, Ref [1].
\vskip .5cm
\noindent
[12] R. Foot and H. Lew, Phys. Rev. D41, 3502 (1990);
Mod. Phys. Lett. A5, 1345 (1990).

\end{document}